\documentclass[twocolumn,secnumarabic,amssymb, nobibnotes, aps, prl]{revtex4-1}

\newcommand{\env}[1]{\texttt{#1}}
\usepackage{bm}
\usepackage{amsmath}
\usepackage{amsthm}
\usepackage{float}
\usepackage{graphics}
\usepackage{graphicx}
\usepackage{MnSymbol}

\usepackage{adjustbox}
\usepackage{subfigure}

\usepackage{adjustbox}
\usepackage{tikz}
\usepackage{grffile}
\usepackage{algorithm}
\usepackage{algpseudocode}

\DeclareMathOperator*{\argmin}{arg\,min}
\usepackage{mathtools} 

\graphicspath{{figures/}}

\usepackage{cleveref}
\crefname{equation}{Eq.}{Eqs.}
\crefname{section}{Sec.}{Secs.}
\crefname{figure}{Fig.}{Figs.}

\newcommand{\be}{\begin{equation}}
	\newcommand{\ee}{\end{equation}} 
\newcommand{\bea}{\begin{eqnarray}}
	\newcommand{\eea}{\end{eqnarray}}

\newcommand{\f}[2]{\frac{#1}{#2}}

\newcommand{\ccup}[1]{\left\{#1\right\}}

\newcommand{\kirk}{Kirchhoff}
\usepackage[normalem]{ulem} 
\usepackage{soul}

\newcommand{\uconstr}{\mbox{{\small Unconstrained}}}

\newcommand{\real}{\mbox{{\small Grenoble}}}

\begin{document}
	
	\title{
		Optimal transport with constraints: from mirror descent to classical mechanics}%
	
	\author{Abdullahi Adinoyi Ibrahim}%
	\email[AAI : ]{abdullahi.ibrahim@tuebingen.mpg.de}
	\affiliation{Max Planck Institute for Intelligent Systems, Cyber Valley, T{\"u}bingen 72076, Germany}
	
	\author{Michael Muehlebach}%
	\email[MM : ]{michaelm@tuebingen.mpg.de}
	\affiliation{Max Planck Institute for Intelligent Systems, Cyber Valley, T{\"u}bingen 72076, Germany}
	
	\author{Caterina De Bacco}%
	\email[CDB : ]{caterina.debacco@tuebingen.mpg.de}
	\affiliation{Max Planck Institute for Intelligent Systems, Cyber Valley, T{\"u}bingen 72076, Germany}
	
	\begin{abstract}
		Finding optimal trajectories for multiple traffic demands in a congested network is a challenging task. Optimal transport theory is a principled approach that has been used successfully to study various transportation problems. Its usage is limited by the lack of principled and flexible ways to incorporate realistic constraints. We propose a principled physics-based approach to impose constraints flexibly in optimal transport problems. Constraints are included in mirror descent dynamics using the principle of D'Alembert-Lagrange from classical mechanics. This results in a sparse, local and linear approximation of the feasible set leading in many cases to closed-form updates. 
	\end{abstract}
		
	\maketitle
	
	\paragraph*{Introduction.} Optimal transport in networks has  important applications in different disciplines, in particular in urban transportation networks \cite{arnott1994economics}.
	Congestion not only increases travel time for users and decreases productivity, but it also drives air pollution. Reducing congestion and making transportation more efficient are also a core objective for EU policies, as highlighted throughout the EU Transport White Paper and the Strategic Plan 2020-2024 \cite{eu2011,eu2021}.
	
	The design of efficient transportation networks is a complex task that requires a multifaceted solution. One of these facets is the problem of finding optimal routes for passengers. This is a well-studied problem in operations research \cite{ahuja1988network} where minimum-cost optimization is often considered to model discrete flows and can be solved using classical techniques from linear programming. In our work, we consider the continuous case, where flows are real-valued quantities.
A variety of approaches have been suggested to model transport in networks using techniques from physics of complex systems \cite{morris2012transport,gao2019effective}.
Path optimality and congestion control have been studied in discrete settings \cite{noh2002stability,dobrin2001minimum,bayati2008statistical} or using the cavity method \cite{yeung2012competition,yeung2013physics}. These usually rely on ad-hoc algorithmic updates that depend on the specific type of constraints. The computational complexity of the ad-hoc updates is greatly influenced by the constraints.
Other approaches have been proposed to investigate navigation in complex systems  \cite{sole2016congestion, gomez2008entropy,lacasa2009jamming,sneppen2005hide,rosvall2005searchability,zhao2005onset,estrada2023network}, where the focus lies on investigating the properties of flows, rather than their optimization, as we consider here. In addition, these models often assume that passengers follow their shortest paths, an assumption, which may not be satisfied in practice.
Adaptation dynamics \cite{tero2010rules, hu2013adaptation, ronellenfitsch2016global} have been proposed to model biological distribution networks. However, these methods fall short of describing realistic scenarios where transport flows are limited by constraints.

In the following we cast the problem of designing efficient transportation networks under the broader framework of optimal transport theory (OT) \cite{santambrogio2015optimal}. This  has been used to
	model and optimize various aspects of transport networks such as network design \cite{tero2010rules,ronellenfitsch2016global, baptista2020network,leite2022revealing} and traffic flows \cite{bonifaci2012physarum,lonardi2021designing,bohn2007structure,ibrahim2021optimal,lonardi2023immiscible}. 
	These approaches guarantee a principled and computationally efficient way of solving transportation problems on networks. In addition, they model traffic congestion with a single tuning parameter that enables a transition between opposite traffic regimes, where traffic congestion can either be consolidated or discouraged.
	In standard OT methods, beyond few obvious constraints (e.g. conservation of mass), the amount of flow passing through an edge of the transportation network is unconstrained. As a result, traffic tends to concentrate on path trajectories that may be structurally unfeasible, which severely limits the applicability of OT models in real-world situations, where, for example, roads have a limited capacity of vehicles traveling at the same time. This letter proposes an approach to avoid this crucial flaw of OT models by imposing constraints. Applying this approach significantly impacts the overall network topology induced by the optimal flows, as the resulting path trajectories have different path lengths and traffic distribution than those obtained from unconstrained scenarios.
	
	Our approach  has not only a solid foundation via the principle of D'Alembert-Lagrange from classical mechanics \cite{lanczos1949variational}, but also leads to algorithms that are computationally efficient and have a low implementation complexity.
	The key idea is to consider mirror descent dynamics of an OT problem, where constraints are included on a velocity level. This leads to a sparse, local and linear approximation of the feasible set which, in many cases, allows for a closed-form update rule, even in situations where the feasible set is nonconvex. 
	\paragraph*{The model.} In analogy with electrical grids or hydraulic networks, we model mass flow on a transportation network using conductivities and flows on network edges. We consider a multi-commodity scenario \cite{lonardi2021designing,bonifaci2022physarum}, where mass of different type $i=1,\dots,M$ can move along different trajectories. The flow $F_e^i$ of mass of type $i$ along an edge $e=(u,v)$ can be described by $F_{e}^{i} = \mu_e (p_{u}^{i} - p_{v}^{i})/\ell_e ,$ where $p_{u}^{i}$ is a pressure potential at node $u$ for passenger of type $i$, $\ell_e$ is the length of the edge $e$ and $\mu_e$ its conductivity. This latter quantity can be seen as proportional to the size of an edge, and is the main variable of interest in determining optimal trajectories. Once the conductivity is known, the pressure differences can then be calculated from \kirk's law, which in turns determines the flows $F_e^i$, see Supporting Material (SM) \cite{suppinfo2023}.\nocite{ibrahim2022sustainable,davis2004algorithm,briggs2000multigrid,nesterov2004introductory} In the absence of constraints, the optimal conductivities are the stationary solutions of the dynamics $\dot{\mu}=f$, where 
	\be\label{eqn:dyn_old} 
	f_{e} = \mu_{e}^{\beta} \f{\sum_{i}(p_{u}^{i} - p_{v}^{i})^{2} }{\ell_{e}^{2}} - \mu_{e} \equiv  \mu_{e}^{\beta- 2} |F_{e}|^{2}- \mu_{e} \quad,
	\ee
	with $F_e= (F_e^1,\dots,F_e^M)$ and $|\cdot|$ denotes the Euclidean norm.
	Intuitively, this equation describes a positive feedback mechanism where conductivities increase for larger fluxes and decrease for negligible ones \cite{tero2010rules}. It can be shown that the dynamics in~\cref{eqn:dyn_old} admits a Lyapunov function $\mathfrak{L}_{\beta_{}}$ which can be interpreted as a combination of the cost to operate the network and that of building the infrastructure \cite{lonardi2021designing}, see SM \cite{suppinfo2023} .  Moreover, we have that $f = -S\, \nabla \mathfrak{L}_{\beta_{}}$, where $S$ is a diagonal matrix with diagonal entries $S_{e}= 2 \mu_{e}^{\beta}/\ell_{e}$ and \cref{eqn:dyn_old} can therefore be seen as a mirror descent for the cost function $\mathfrak{L}_{\beta_{}}$ \cite{bonifaci2021laplacian}. This scaling in $S$ has the advantage of ensuring good behavior of the resulting numerical methods. One can also reinterpret \cref{eqn:dyn_old} as a classical gradient descent by applying a suitable transformation \cite{facca2021fast},  we do not explore this here.
	
	Variants of these dynamics have been proposed to model distributions over networks \cite{hu2013adaptation, bohn2007structure, katifori2010damage, banavar2000topology,ronellenfitsch2016global}.
	The constant $\beta \in (0,2)$ regulates the desired transportation regime. The setting $\beta<1$ penalizes traffic congestion by distributing paths on more edges, $\beta>1$ encourages path consolidation into fewer highways, and $\beta=1$ is shortest path-like.
	
	In addition to imposing \kirk's law on nodes to ensure mass conservation, solving these dynamics outputs otherwise unconstrained optimal $\mu_{e}$ and $F_e$ (see SM \cite{suppinfo2023}). While this may be enough in ideal cases, in more realistic scenarios it is important to further constrain the solution. For instance, structural constraints may limit the maximum amount of flow that an edge can carry, or a budget constraint may be used to limit the infrastructure cost for building the network. Hence, the dynamics $\dot{\mu}=f$ must be altered to account for these additional constraints.\\
	There are many ways in which constraints can be added. A popular approach is to add constraints on a so-called position level, which leads to gradient inclusions in continuous time \cite[Ch~3.4]{AubinCellina}, and projected gradient descent in discrete time. Unfortunately, the scope of projected gradients is limited, due to the fact that projections can only be efficiently evaluated for constraints that have a particular structure (such as a low-dimensional hyperplane, the probability simplex, or a Euclidean norm ball). When the feasible set is nonconvex and/or fails to have a simple structure, evaluating projections is a computationally daunting task. This motivates our formulation (see also \cite{muehlebach2021constraints}), which includes constraints on a velocity level and yields a sparse local and linear approximation of the feasible set. As a consequence, the updates for $\mu$ can often still be evaluated in closed-form (or there is an efficient way of computing them numerically) even though the underlying feasible set is nonconvex or fails to have a simple structure. We will highlight explicit examples of such situations in the remainder of this letter. 
	
	We define $C := \{ \mu_{} \in \mathbb{R}_{\geq0}^{E} \ | \ g(\mu_{}) \ge 0 \} $ as the set of feasible conductivities $\mu_{}=(\mu_{1},\dots,\mu_{E})$, with $g$ a constraint function that we assume continuously differentiable and $E$ is the number of network edges.
	Interpreting $\mu$ as a “position” variable  we can equivalently express the constraints in $C$ in terms of a ``velocity'' variable by imposing $\dot{\mu}(t)\in V_\alpha(\mu(t))$, where 
	$V_\alpha(\mu(t))$ is the set of feasible velocities and $\alpha \geq 0$ is a constant typically referred to as a ``restitution'' parameter or ``slackness'',  see Appendix for details.

\begin{figure}
		\centering
		\includegraphics[width=0.99\linewidth]{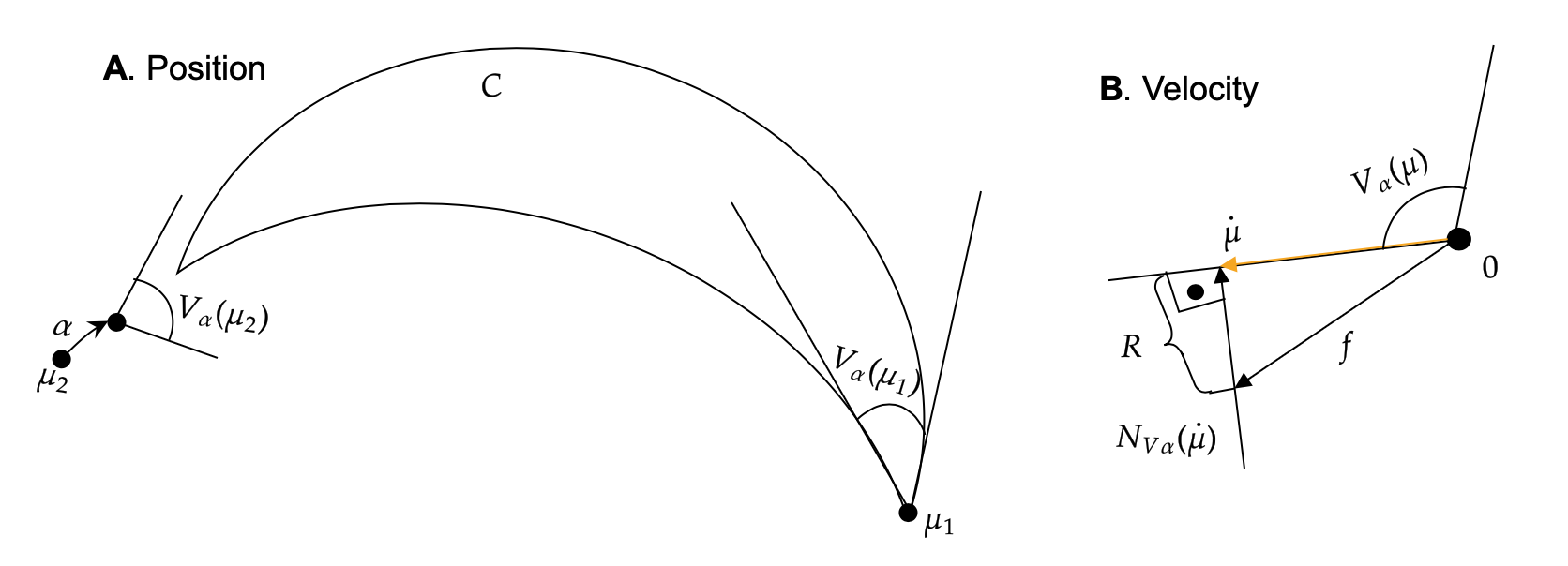}
		\caption{(A) Visualization of the set $C$ and the set of feasible velocities $V_{\alpha}(\mu_{1})$ and $V_{\alpha}(\mu_{2})$ at points $\mu_{1}$ and $\mu_{2}$, respectively. Point $\mu_{1}$ lies on the boundary of $C$, while $\mu_{2}$ is infeasible; $\alpha$ is a restitution parameter. (B) When the vector field $f$ is pushing away from $C$, a force $-R\in N_{V_{\alpha}}(\dot{\mu}_{})$ is added to the dynamics to ensure $\dot{\mu}\in V_{\alpha}(\mu)$ .} 
	\label{fig:setC}
\end{figure}

	For $\mu(t)\not\in C$ and an active constraint $i$, the constraint $\dot{\mu}(t)\in V_\alpha(\mu(t))$ is equivalent to $\text{d}g_i(\mu(t))/\text{d}t\geq - \alpha g_i(\mu(t))$, which ensures that potential constraint violations decay at the rate $\alpha>0$. The situation is visualized in \Cref{fig:setC}(A).
	
In order to account for the velocity constraint $\dot{\mu}\in V_\alpha(\mu)$ we augment the dynamics $\dot{\mu}=f$ with a \textit{reaction} force $R$ that forces the solution to remain within the desired constraints:
\be\label{eqn:new-dyn-withR}
\dot{\mu} = f + R,  \quad \text{with}\,\ - R \in  N_{V_{\alpha} (\mu)}(\dot{\mu}),
\ee
where $N_{V_\alpha(\mu)}(\dot{\mu})$ denotes the normal cone of the set $V_\alpha(\mu)$ at $\dot{\mu}$. Due to the scaling of the gradient with $S$, the normal cone is defined with respect to the inner product $\langle a,b\rangle=a^{T} S^{-1} b$, where $a,b\in \mathbb{R}^E$ are arbitrary vectors. This has the important effect of guaranteeing that $\mathfrak{L}_{\beta_{}}$ (of the unconstrained dynamics) is still a Lyapunov function also in the constrained setting and that $\mathfrak{L}_{\beta_{}}(\mu(t))$ is monotonically decreasing along the trajectories of \cref{eqn:new-dyn-withR}. 
A detailed derivation is included in SM \cite{suppinfo2023}.

The addition of $R$ ensures that even if $f$ pushes $\mu$ away from $C$, as shown in \Cref{fig:setC}(B), the force $R$, which is orthogonal to the set $V_\alpha(\mu)$, annihilates the component of $f$ that would lead to a constraint violation and ensures that $\dot{\mu} \in V_\alpha(\mu)$. As discussed above, we can therefore conclude that $\mu(0)\in C \Rightarrow \mu(t)\in C$ for all $t\geq 0$ and $\mu(0)\not\in C \Rightarrow \mu(t)\rightarrow C$ for $t\rightarrow \infty$. 

In addition, we infer from \Cref{fig:setC} that the resulting $\dot{\mu}$ in \cref{eqn:new-dyn-withR} is nothing but the projection of $f$ onto the set $V_\alpha(\mu)$ and as a result, we can rewrite $\dot{\mu}$ in the following way:
\be\label{eqn:updated_dyn}
\dot{\mu}_{}  := \argmin_{v \in  V_{\alpha}(\mu_{})  } 
\f{1}{2} \langle v-f, v-f \rangle \quad,
\ee
which can also be equivalently reformulated as the quadratic program (QP)
\begin{align}\label{eqn:updated_dyn3}
	\dot{\mu}_{}  &:= \argmin_{v \in  V_{\alpha}(\mu_{})  } 
	\f{1}{2} (v-f)^{T} S^{-1} (v - f) \quad.
\end{align}

This reformulation is not only useful for numerical computations, but also highlights that the velocity $\dot{\mu}$ is chosen, at each point in time, to match
the unconstrained $f$. \Cref{fig:setC}(A) visualizes the set $C$ and the set of feasible velocities $V_{\alpha}(\mu_{1})$ and $V_{\alpha}(\mu_{2})$ at points $\mu_{1}$ and $\mu_{2}$, respectively. Point $\mu_{1}$ lies on the boundary of $C$, while $\mu_{2}$ is infeasible. We note that the cone $V_\alpha(\mu_{2})$ includes an offset, which is controlled by the restitution parameter $\alpha$; this ensures that any $v\in V_\alpha(\mu_{2})$ leads to a decrease in constraint violation. \cref{fig:setC} (B) shows that when the vector field $f$ is pushing away from $C$, a force $-R\in N_{V_{\alpha}}(\dot{\mu}_{})$ is added to the dynamics. The force $R$ annihilates the component of $f$ that would lead to a constraint violation and ensures $\dot{\mu}\in V_{\alpha}(\mu)$, where $\dot{\mu}$ is chosen as close as possible to $f$. This can also be interpreted as Gauss's principle of least constraint. It is important to note that $V_\alpha(\mu)$ is a polyhedral set that only includes the constraints $I_\mu$, a subset of the original constraints $g(\mu)\geq 0$. The set $V_\alpha(\mu)$ represents therefore a sparse, local and linear approximation of the feasible set. The solution $\dot{\mu}$ of \cref{eqn:updated_dyn} can then be used to update the conductivity with a discrete-time algorithm:
\be\label{eqn:update}
\mu^{t+1} = \mu^{t} + \tau \dot{\mu}\quad,
\ee
where $\tau>0$ is the step size.

This general formalism can be applied to a variety of scenarios, provided one can compute $\nabla g$, which determines the set $V_{\alpha}(\mu)$. We can then solve \cref{eqn:updated_dyn3} by using numerical solvers tailored to QP, which then yields the update \cref{eqn:update}. Additional details about the computational complexity for solving \cref{eqn:update} are described in SM \cite{suppinfo2023}. However, in important special cases, the optimization \cref{eqn:update} can be solved in closed-form, as we illustrate below with three relevant examples.

\paragraph*{Capacity constraints.} In cases of structural constraints that strictly limit the amount of mass that can travel along any given edge, one can consider capacities $c_e \geq 0$ on edges and set  constraints as $g_e(\mu) = c_e - \mu_{e}$. 
The velocity constraint $v \in V_\alpha(\mu)$ in \cref{eqn:updated_dyn} reads as $v_e\le \alpha g_{e}(\mu_e)$, for $e\in I_{\mu}$, which is strictly negative, since $\alpha >0$ (SM \cite{suppinfo2023}). 
As previously discussed, $\alpha > 0$ is a restitution parameter that dictates the rate at which constraint violations decay. In discrete time, one should choose $\alpha>0$ such that $\alpha\, \tau \leq 1$ to guarantee convergence (see \cite{muehlebach2021constraints}).
We can then solve \cref{eqn:updated_dyn} in closed-form for edges violating the constraint obtaining $v_e = \min \ccup{ \alpha \,(c_{e}-\mu_{e}), {f}_{e}}$.
In summary, for each edge $e$, we have:
\be
\dot{\mu}_{e} =
\begin{cases}
	\alpha \,(c_{e} - \mu_{e}), & \text{if} \ \	{f}_{e} \ge  \alpha\,(c_{e} -  \mu_{e})\ \text{and}\ \mu_{e}\ge c_e,\\\
	{f}_{e} &  \text{otherwise}\quad .
\end{cases}
\ee

\begin{figure}
	\includegraphics[width=1.04\linewidth]{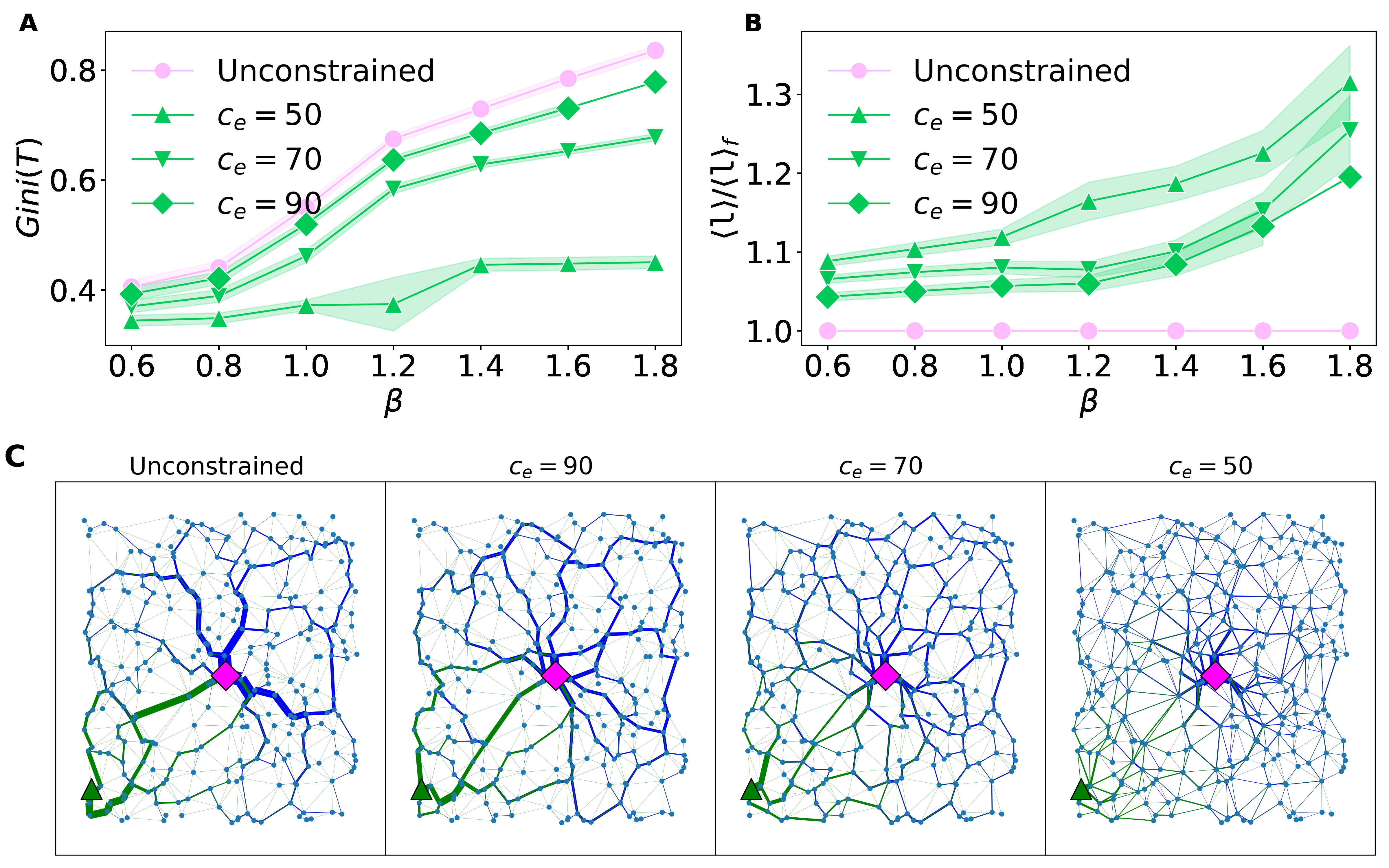}
	\caption{Capacity constraint on synthetic networks. (A) Gini coefficient of the traffic distribution on edges. The edge capacity $c_e=c$ is selected as a percentile of the distribution of $\mu$ over edges obtained in the unconstrained case (\uconstr{}). (B) Ratio of average total path length to that of \uconstr{}, $\langle \mathit{\mathtt{l}} \rangle_{f}$.  Markers and shadows are averages and standard deviations over 20 network realizations, with 100 randomly selected origins. All passengers have the same central destination (square magenta marker). (c) Example trajectory of one passenger type (green color), whose origin is the green triangle marker. Edge widths are proportional to the amount of passengers traveling through an edge; $\beta=1.8$. }
	\label{fig:capacity_results}
\end{figure}

\Cref{fig:capacity_results} shows the path topologies with capacity constraints on synthetic data, compared against the unconstrained case. We generate random planar networks as the  Delaunay triangulation \citep{guibas1985primitives} of $N=300$ points
in the plane.
We measure the Gini coefficient $Gini(T)$ calculated on the traffic on edges, defined as the $E$-dimensional vector $T$ with entries $T_e = \sum_i |F_e^i|/n$, where $n$ is the number of passengers. The coefficient has value in $[0,1]$ and it determines how traffic is distributed along network edges, with $Gini(T)=0,1$ meaning equally-balanced or highly unbalanced traffic on few edges, respectively. The choice of the edge capacity $c_{e}$ influences this value, with lower $c_{e}$ imposing stricter constraint and thus encouraging traffic to distribute more equally along the edge, i.e. lower Gini, as shown in \cref{fig:capacity_results}(A). Conversely, this implies longer routes for passengers, as measured by an increasing average total path length $\langle l \rangle = \sum_{e,i}\, \ell_e\, |F_e^i|/n$  compared to the unconstrained solution, as shown in \cref{fig:capacity_results}(B).

\paragraph*{Budget constraint.}
As a second example, we consider a global constraint that involves all the edges at once, a budget constraint $g_{b}(\mu) = b\,-\,\sum_{e}\mu_{e}^{}$. This is relevant when a network manager has a fixed limited amount of resources $b>0$ to invest. 
We note that, while the Lyapunov function $\mathfrak{L}_{\beta_{}}$ contains a similar budget term--the cost to build the infrastracture--this cost is not regarded as a constraint
in standard approaches \cite{hu2013adaptation,lonardi2021designing} but as part of the energy consumption, and the budget $b$ is not a Lagrange multiplier but a measurable constant. Furthermore, unlike the previous case where including a positivity constraint $\mu_e \ge 0$ is optional (but it can in principle be imposed as well, see SM \cite{suppinfo2023}), here we need to include that explicitly. In the standard OT formalism positivity is ensured, provided $\mu_e$ is initialized as a positive quantity. Adding constraint may not preserve positivity anymore during the updates, this is the case for the budget constraint, as we observed empirically. Positivity is enforced by adding $g_{p}(\mu) = \mu \geq 0$, i.e. $\mu_{e}\ge 0\, \forall e$.

In this budget constraint setting, the conductivities violate the constraint whenever $\sum_{e} \mu_{e} > b$. 
We derive a closed-form solution as: $\dot{\mu}_{e} = f_{e} -  S_{e}^{} \lambda_b$, if $f_{e} -  S_{e}^{} \lambda_b \ge -\alpha\, \mu_{e}$, and $\dot{\mu}_{e} = -\alpha \,\mu_{e}$ otherwise, where $\lambda_b \in \mathbb{R}$, a Lagrange multiplier for the budget constraint, can be numerically determined via fixed-point iteration (SM \cite{suppinfo2023}).

\paragraph*{Combining linear and non-linear constraints.} All the previous examples considered linear constraints, where it is simple to derive analytical solutions. In general, constraints can be more complicated and thus require numerical methods to solve the constrained QP in \cref{eqn:updated_dyn}.
In this scenario, we consider a non-linear budget constraint of the form: $g_{\delta}(\mu) = b-\sum_{e}\mu_{e}^{\delta}\geq 0$, where $\delta >0$ is a nonlinearity parameter. Setting $\delta=1$ gives a linear budget constraint as the one discussed earlier. A non-linear example is a volume-preserving constraint where $\delta = 1/2$, this is relevant for biological processes such as leaf venation and  vascular systems \cite{takamatsu2017energy,ronellenfitsch2016global}. This non-linear budget induces the velocity constraint $\sum_{e} \delta \mu_{e}^{\delta -1}v_{e} \le \alpha \, g_{\delta}(\mu)$.
In addition, we also consider a capacity constraint as in the first scenario studied above.
Overall, three functions are required: i) $g_{\delta}(\mu)$ to impose non-linear budget constraint; ii) $g_{e}(\mu)$ to impose edge capacity and iii) $g_{p}(\mu)$ to ensure positivity. 
We derive the closed-form solution as\be
\dot{\mu}_{e} =
\begin{cases}
	\alpha\,(c_{e} - \mu_{e}) & \text{if}  \	{f}_{e}{- S_{e}^{} \lambda_{\delta}\,h_e} \ge  \alpha\, (c_{e} - \mu_{e}), \, \mu_e\geq c_e 
	\\\\ 
	- \alpha \,\mu_{e} &  \text{if} \  {f}_{e} {- S_{e}^{} \lambda_{\delta}\, h_e} \le -\alpha\, \mu_{e}, \,\mu_e \leq 0
	\\\\ 
	f_{e} - S_{e}^{} \lambda_{\delta}\,h_e &  \text{otherwise} \quad, 
\end{cases}
\ee 
where $h_{e} = \delta \,{\mu}_{e}^{\delta - 1}$ and $\lambda_{\delta} > 0$. The value of $\lambda_{\delta}$ can be determined numerically using fixed-point iteration (SM \cite{suppinfo2023}). The value $ \alpha\,(c_{e}-\mu_{e})$ ensures there is no violation on the edge capacity, $ -\alpha\,\mu_{e}$ imposes positivity constraint and ${f}_{e} {- S_{e}^{} \lambda^{\delta} h_e}$ captures budget violation.  Overall, this scenario ensures that the velocity $\dot{\mu}_{e}$ has an upper bound of $\alpha\,(c_{e}-\mu_{e})$ and lower bound of $-\alpha\,\mu_{e}$. The choice of $\delta$ impacts the topological properties of the resulting network, e.g., the total path length. In the numerical experiments, we set the nonlinearity parameter as $\delta \in  (0,1)$.

\begin{figure*}
	\centering
	\includegraphics[width=1\linewidth]{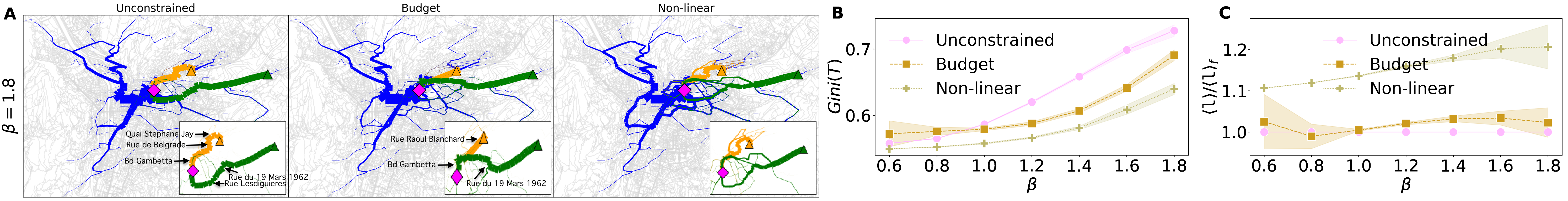}
	\caption{ Constrained OT on \real{} road network. (A) Path trajectories for the unconstrained OT (\uconstr{}), budget constraint (Budget) and a non-linear budget plus capacity (Non-linear). We set $b = \frac{1}{2}\sum_{e} \mu_{e}$, where $\mu_{e}$ is that of unconstrained, $\delta=1/2$ and $c_{e} = 70$ for all edges. Example trajectories of two passenger types (green and orange), whose origin are the respective triangles. All passengers have the same central destination (magenta marker). Edge widths are proportional to the amount of passengers traveling through an edge. (B) Gini coefficient
		of the traffic distribution on edges. (C) Ratio of average total path length to that of Unconstrained. Markers and shadows
		are averages and standard deviations over 100 randomly-selected destinations, respectively. }
	\label{fig:real_topology}
\end{figure*}

\paragraph*{\real{} network.} We examine the topology of various constrained solutions on the road network of the city of \real{} \cite{kujala2018collection}, see \cref{fig:real_topology}(A). This has 640 nodes and 740 edges. As a relevant example, we set the central bus station as the destination node and select the remaining $639$ nodes as origins, but our method still applies to other choices of origin-destination pairs, e.g. peripheral nodes connecting to other peripheral nodes or to various hubs. This can be specified inside Kirchhoff’s law, see SM \cite{suppinfo2023}.

Routes generated from the non-linear constraint scenario balance traffic more than the unconstrained case and result in longer routes, see \cref{fig:real_topology}(B-C). 
Adding a budget constraint for $\beta>1$ results in more distributed traffic (lower Gini) without increasing much the total path length, compared to the unconstrained case. This could be used for instance to allocate  to roads infrastructural works aimed at maintenance or upgrade when having a restricted budget.

\paragraph*{Discussion.}
Distributing flows in a transportation network is challenging. Approaches based on optimal transport theory are promising, but they are limited by the lack of a mechanism to incorporate realistic constraints.
We show how to impose arbitrary constraints on OT problems in a principled and flexible way. The constraints are lifted from a position to a velocity level and are included in the corresponding mirror descent dynamics. This results in a scalable algorithm that solves constrained OT problems in a computationally efficient manner. The algorithm relies on a sparse local approximation of the feasible set at each iteration. Thus, closed-form updates can often be derived, even if the underlying feasible set is nonconvex or nonlinear. Otherwise, one can resort to efficient numerical methods to solve at most a quadratic program. Our physics-based approach is a change of paradigm with regard to how OT problems are modelled and solved numerically. 
This calls for a generalization of transportation problems in wider scenarios, e.g. in networks with multiple transport modes \cite{ibrahim2021optimal}, with real-time traffic demands \cite{lonardi2023infrastructure} or with noise-induced resonances \cite{folz2022noise}.

We provide an open source implementation \cite{mcopt23}.
\\ 

\begin{acknowledgments}
	\env{Acknowledgments:}	The authors thank the International Max Planck Research
	School for Intelligent Systems (IMPRS-IS) for supporting AAI. MM thanks the German Research Foundation and the Branco Weiss Fellowship, administered by ETH Zurich, for the support.
\end{acknowledgments}

\newpage

\appendix
\section{Appendix}\label{apx:constraints}

\paragraph*{Details about setting the constraints.}	We define $C := \{ \mu_{} \in \mathbb{R}_{\geq0}^{E} \ | \ g(\mu_{}) \ge 0 \} $ as the set of feasible conductivities $\mu_{}=(\mu_{1},\dots,\mu_{E})$, with $g$ a constraint function that we assume continuously differentiable and $E$ is the number of network edges.
We focus on those edges where  constraints are not satisfied, and denote the set of active constraints for a given $\mu$ as $I_{\mu_{}}:= \{i \in \mathbb{Z} \ | \ g_{i}(\mu_{}) \le 0 \}$. Interpreting $\mu$ as a “position” variable, a constraint to ensure $\mu(t) \in C, \forall t\geq 0$, can be equivalently formulated as a constraint on its \textit{velocity} $\dot{\mu}(t)\in T_{C}(\mu(t)), \forall t\geq 0$, with $\mu(0)\in C$, where $T_{C}(\mu)$ denotes the tangent cone of the feasible set at $\mu$, see \cite{rockafellar1998variational}. However, it will be convenient to slightly extend the notion of tangent cone to also account for infeasible initial conditions (this is particularly important for the discretization), which is achieved by imposing $\dot{\mu}(t)\in V_\alpha(\mu(t))$, where 
	$V_{\alpha}(\mu_{}) := \{ 
	v\in \mathbb{R}^{E} \ | \ \nabla g_{i}(\mu_{})^{T} v \ge -\alpha\, g_{i}(\mu), i \in I_{\mu_{}}
	\}$, and $\alpha \geq 0$ is a constant typically referred to as a ``restitution'' parameter or ``slackness''.
	We note that $V_\alpha(\mu)$ generalizes the notion of the tangent cone, since for $\mu\in C$, $V_\alpha(\mu)=T_C(\mu)$. We assume mild regularity conditions (constraint qualification). A sufficient condition is, for example, the existence of $v\in \mathbb{R}^E$ such that $\nabla g_i(\mu)^T v > 0$ for all $i\in I_\mu$. 
	
For $\mu(t)\not\in C$, the constraint $\dot{\mu}(t)\in V_\alpha(\mu(t))$ is equivalent to $\text{d}g_i(\mu(t))/\text{d}t\geq - \alpha g_i(\mu(t))$, $i\in I_{\mu(t)}$, which ensures that potential constraint violations decay at the rate $\alpha>0$.
	
\paragraph*{Details about our method.} From a variational optimization perspective, our approach is related to successive linear and sequential quadratic programming \cite{nocedal2006,bertsekas1999,luenberger2016}. The underlying idea of these methods is to linearize the objective function and the constraints about the current iterate and to solve a local linear and/or quadratic program. Our work improves upon these ideas and tailors these to optimal transport problems in the following way: i) we linearize a subset of constraints at every iteration, which means that the subproblem \Cref{eqn:updated_dyn} typically includes very few constraints and can be solved efficiently; ii) we introduce a non-Euclidean inner product that is adapted to optimal transport problems and is used to show that $\mathfrak{L}_{\beta_{}}$ is a Lyapunov function; iii) we provide closed-form updates in various problem instances that are practically relevant.

\clearpage 
\bibliography{bibliography}

\begin{thebibliography}{52}
\bibitem [{sup()}]{suppinfo2023}%
  \BibitemOpen
  \href@noop {} {\bibinfo  {journal} {See Supporting Material, which includes
  Refs. [27-31], for additional information about the derivations and a
  detailed discussion of the numerical simulations}\ }\BibitemShut {NoStop}%
\end{thebibliography}

\end{document}